\documentclass{pasj00}

\begin{document}
\SetRunningHead{T. Wada et al.}{The AKARI/IRC deep survey}
\Received{2007/06/03}
\Accepted{2007/08/14}

\title{The Infrared Camera (IRC) deep survey in the performance verification phase}

\author{
   Takehiko \textsc{Wada},\altaffilmark{1}
    \thanks{Further information contact Takehiko Wada
    (wada@ir.isas.jaxa.jp)}
   Shinki \textsc{Oyabu},\altaffilmark{1}
   Yoshifusa \textsc{Ita},\altaffilmark{1}
   Hideo \textsc{Matsuhara},\altaffilmark{1}
   Chris P. \textsc{Pearson},\altaffilmark{1,2}
   Takashi \textsc{Onaka},\altaffilmark{3}
   Youichi \textsc{Ohyama},\altaffilmark{1}
   Fumihiko \textsc{Usui},\altaffilmark{1}
   Naofumi \textsc{Fujishiro},\altaffilmark{1}
   Daisuke \textsc{Ishihara},\altaffilmark{3} 
   Hirokazu \textsc{Kataza},\altaffilmark{1}
   Woojung \textsc{Kim},\altaffilmark{1}
   Toshio \textsc{Matsumoto},\altaffilmark{1}
   Hiroshi \textsc{Murakami},\altaffilmark{1}
   Itsuki \textsc{Sakon},\altaffilmark{3} 
   Toshihiko \textsc{Tanab\'{e}},\altaffilmark{4}
   Toshinobu \textsc{Takagi},\altaffilmark{1}
   Kazunori \textsc{Uemizu},\altaffilmark{1}
   Munetaka \textsc{Ueno},\altaffilmark{5}
   and
   Hidenori \textsc{Watarai}\altaffilmark{6}
}

 \altaffiltext{1}{Institute of Space and Astronautical Science, Japan Aerospace Exploration Agency,
   Sagamihara, Kanagawa 229 8510 } 
 \altaffiltext{2}{ISO Data Centre, ESA, Villafranca del Castillo, Madrid, Spain.}
 \altaffiltext{3}{Department of Astronomy, School of Science, University of Tokyo, Bunkyo-ku, Tokyo 113-0033 }
 \altaffiltext{4}{Institute of Astronomy, University of Tokyo, Mitaka,Tokyo 181-0015}
 \altaffiltext{5}{Department of Earth Science and Astronomy, Graduate
   School of Arts and Sciences,
 The University of Tokyo, Meguro-ku,Tokyo 153-8902}
 \altaffiltext{6}{Office of Space Applications, Japan Aerospace Exploration Agency, Tsukuba, Ibaraki 305-8505 }


\maketitle

\begin{abstract}
We report the first results of a near- and mid- infrared deep survey
with the Infrared Camera (IRC) onboard AKARI 
in the performance verification phase.
Simultaneous observations by the NIR, MIR-S and MIR-L channels of the IRC
with effective integration times of 4529, 4908, and 4417 seconds at 3, 7, and 15 $\mu$m, covering 86.0, 70.3, and 77.3
arcmin$^2$ area, detected 955, 298 and 277 sources, respectively.
The 5 $\sigma$ detection limits of the survey are
6.0, 31.5 and 71.2$\mu$Jy and the 50 \% completeness limit 
are  24.0, 47.5, and  88.1$\mu$Jy at 3, 7, and 15 $\mu$m, respectively.
The observation is limited by source confusion at 3 $\mu$m.
We have confirmed the turnover in the 15 $\mu$m differential source counts 
around 400 $\mu$Jy, previously detected by surveys with the Infrared 
Space Observatory.
The faint end of 15 $\mu$m raw source counts agree with the results 
from the deep surveys in the GOODS fields  carried out with the {\it Spitzer} IRS peak up imager 
and the predictions of current galaxy evolution models.
These results indicate that deep surveys with comprehensive
 wavelength coverage at mid-infrared wavelength are very important to
 investigate the evolution of infrared galaxies at high redshifts.
\end{abstract}

\section{Introduction}
The Infrared Camera (IRC; \cite{onaka-irc-inst};
\cite{wada-irc-cospar})
is a near- and mid- infrared scientific instrument onboard 
the first dedicated Japanese infrared astronomical satellite AKARI
 (\cite{murakami-astrof}).
The unique instrumental features of the IRC, compared to the similar
contemporary IRAC instrument on the  {\it Spitzer} Space Telescope
\citep{fazio-irac}, 
are a wider field of view (FOV; 10 $\times$ 10 arcmin$^2$) 
and a more comprehensive contiguous wavelength coverage (2-26$\mu$m) 
with a larger numbers of filters (9 bands)  of comparable sensitivity to
IRAC.
A deep and wide field near- and mid- infrared survey (LS NEP DEEP) 
covering these wavelengths by all the 9 bands 
is underway in the North Ecliptic Pole (NEP) region
\citep{matsuhara-lsnep}.

Previous deep surveys at mid-infrared wavelengths have made 
the striking discovery of a bump in the Euclidean-normalized
differential source counts around 0.4 mJy at 15 $\mu$m 
in the ISO/ISOCAM surveys \citep{elbaz-iso15} 
and 0.3 mJy at 24 $\mu$m in the {\it Spitzer} MIPS surveys \citep{papovich-mips24}.
\citet{marleau-fls24} compared the 15 $\mu$m and 24 $\mu$ source counts 
adopting a 24/15 $\mu$m flux ratio of 1.2, which was derived from the SED
template of \citet{chary-dusty-sfr}. They found that a shift in the
turnover of the Euclidean-normalized counts, however, the fainter side of
15 $\mu$m counts was not well constrained due to the  limitation of the
sensitivity and areal coverage of the ISO surveys.
\citet{pearson-sst-iso-model} showed that models incorporating 
either a burst of evolution or a more smoother continuous
evolution could simultaneously reproduce 
both the 15 $\mu$m ISO and 24 $\mu$m {\it
Spitzer} counts and concluded that 
revealing the dominant population at the fainter end of the 15
$\mu$m counts (e.g., starburst, LIRGs, ULIRGs) was 
an important point to be resolved.

One of the objectives of the LS NEP DEEP survey is to reveal the nature of
the faint 15 $\mu$m population through a deep and wide multi-band survey
together with the SED fitting technique focused on the PAH emission 
and the silicate absorption features \citep{takagi-silicate-break}.
Note that although {\it Spitzer} also has the capability,
 via its IRS 16$\mu$m peak-up Imaging (PUI) mode, for imaging in this
 wavelength range (\cite{teplitz-16um-goods-n};
 \cite{kasliwal-16um-bootes}), however, 
the FOV of the 16$\mu$m PUI mode is very small ($1^{'} \times 1^{'}$; 100 times
smaller compared to that of the IRC), 
and moreover, there still exists a gap in the wavelength
coverage between the IRS 16$\mu$m PUI and the shorter 8$\mu$m IRAC band.

In order to evaluate the capability of such a deep and wide survey by 
the IRC, we have conducted a pilot survey in both the imaging and
spectroscopic modes with 10 pointed observations for each mode
at the end of the performance verification phase of the AKARI satellite.

In this paper, we report on the first results of the pilot imaging survey.
The result of the corresponding pilot spectroscopy survey will be reported elsewhere.
In this work, section \ref{sec-observation} describes the observations.
In section  \ref{sec-reduction}, the data
reduction is presented,
source extraction and photometry are described in section
\ref{sec-catalog}, section \ref{sec-sensitivity} describes the estimation of sensitivity and
completeness of the survey.
Finally, we discuss the capability of the IRC in  deep survey mode
and the 15 $\mu$m source counts in section \ref{sec-discussion}.

\section{Observation}
\label{sec-observation}

\subsection{AOT for deep surveys}

Since the main objective of the AKARI mission is the far-infrared
All-Sky Survey \citep{pearson-ass}, the orbit is by design a
Sun-synchronous polar orbit which puts severe constraints on the
visibility of individual target fields on the sky. Thus,  the operation
for any pointed observations have been made 
as simple as possible, producing  strong constrains
 on the Astronomical
Observation Templates (AOT) for the IRC pointed observations such as (1)
fixed frame integration time, (2) no mapping mode in a single pointed
observation (approximately 10 minutes), that were not apparent for
infrared space observatories such as {\it Spitzer} or the Infrared Space
Observatory (ISO)\footnote{
These constraints make the IRC 
inefficient for wide shallow surveys 
in which a depth of one pointed observation or better
is not required. We have prepared a separate IRC scanning mode
for such a shallow and wide surveys including the IRC All-Sky Survey
at mid-infrared wavelengths \citep{ishihara-scan}.}.

A series of three AOT sets for pointed observations with the IRC have
been prepared.
The first set, referred to as IRC02 and IRC03, are for medium deep 
multi-band surveys,
in which both changes of filters and dithering in target position
is performed.
The second set are for deep surveys and are referred to as IRC00 and IRC05,
in which no filter change and no dithering is performed,
in order to minimize the dead time and maximize the observation time.
For the IRC00 and IRC05 AOTs, at least three pointed observations are
required in order to ensure redundant and reliable observations.
The third AOT set, referred to as IRC04 is designed for spectroscopic
observations and is described elsewhere  \citep{ohyama-irc-spec}.

Using the AOTs designed for deep surveys, a deep and wide field near- and mid- infrared survey (LS NEP DEEP) 
is underway in the North Ecliptic Pole (NEP) region \citep{matsuhara-lsnep}.
In order to evaluate the capability of deep imaging surveys by 
the IRC, we have conducted a pilot survey with 10 pointed observations.

\subsection{pilot survey}
The observations were executed from April 29 to May 6 in 2006, 
almost at the end of the performance verification (PV) phase of AKARI.
We selected a target position of (alpha, delta)=(268.88, 66.61),
chosen by the following reasons.
(1) The FOV of the NIR and MIR-S channels must be within the continuous
viewing zone of AKARI (CVZ: within 0.6 degree from the North or South ecliptic poles).
(2) The field must be well studied by the other observations
in order to make cross-identification of the AKARI sources.
The deep multi-color optical images (\cite{wada-scam}) taken by
Subaru/Suprime-Cam (\cite{miyazaki-scam}) were used for this purpose.
(3) The field must have few bright objects in order to make the deep survey.
We carefully selected a field, which had no bright stars, in the
Suprime-Cam image field.

We used the AOT for deep observation (IRC00) with an AOT filter combination
parameter of ``b'', which corresponds to the filter set of 
the N3 (3.3 $\mu$m), S7 (7.0 $\mu$m) and L15 (15.0 $\mu$m) band filters, 
for the NIR, MIR-S and MIR-L channels, respectively.
We used a target position parameter of ``N'', which means
that the center of the FOV of the NIR channel
(and simultaneously, the MIR-S channel which observes the same point on
the sky  via a beam splitter)
targeted the specified coordinates, while that of the MIR-L channel
was pointed at a parallel field, around 25 arcmin apart from that of
the NIR/MIR-S.
No dithering operation between pointed observations were made,
i.e. we used the same target position for all pointed observations\footnote{
Dithering operation between pointed observations have already implemented 
in the observations after 2006-09-24T09:00:00}.
However, the FOV of the NIR/MIR-S rotates around the center of the field
as the direction of the Sun changes (approximately one degree per day),
resulting in good redundancy except at 
the very center of the FOV.
Note that good redundancy in the observations is possible over the entire FOV of the MIR-L channel,
since the FOV of the MIR-L rotates around the FOV of NIR/MIR-S.
Details of the observation parameters are summarized in Table~\ref{tab:obslog}.

Typically, 10 and 30 frames with effective exposure times 
of 44.4 and 16.4 seconds in each frame were obtained for a single pointed observation, 
for the NIR and MIR-S/L channels, respectively.
The total exposure time was 4530, and 4908 and 4417 seconds, for the NIR,
MIR-S and MIR-L channels, respectively.

\begin{figure}
   \begin{center}
      \FigureFile(80mm,50mm){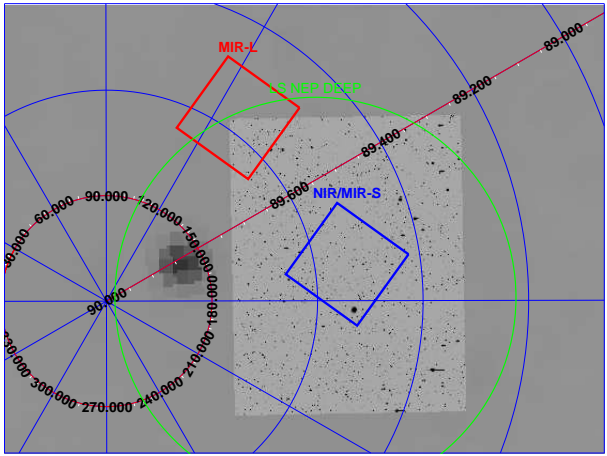}
   \end{center}
   \caption{
The locations of the survey fields and the SUBARU/Suprime-Cam 
deep z' band image, superposed on IRAS 12 micron map.
Blue box and red box are 10$\times$10 arcmin field of view
of the NIR/MIR-S and MIR-L channels, respectively.
The N3 and  S7 fields are covered by the Suprime-Cam image, 
while the L15 field is partially covered by the Suprime-Cam image.
The axes shown in the figure are in ecliptic coordinate.
The field of the AKARI large area deep survey (LS NEP DEEP) 
is indicated as a green circle.
}\label{fig:fov}
\end{figure}

\begin{table*}
  \caption{Observation log}\label{tab:obslog}
  \begin{center}
    \begin{tabular}{lllllll}
      \hline
      Target id & AOT & OBS-DATE &
     \multicolumn{2}{c}{N/S position} & \multicolumn{2}{c}{L position}\\
          &           &                     & RA & DEC & RA &DEC\\
5020053.1 & IRC00 b;N & 2006-05-03T03:32:41 & 268.88 & 66.61 & 269.31 & 66.90\\
5020053.2 & IRC00 b;N & 2006-05-03T06:50:31 & 268.88 & 66.61 & 269.31 & 66.90\\
5020053.3 & IRC00 b;N & 2006-05-03T20:02:41 & 268.88 & 66.61 & 269.30 & 66.90\\
5020053.4 & IRC00 b;N & 2006-05-03T21:41:17 & 268.88 & 66.61 & 269.30 & 66.90\\
5020053.5 & IRC00 b;N & 2006-05-03T23:21:02 & 268.88 & 66.61 & 269.30 & 66.90\\
5020054.1 & IRC00 b;N & 2006-04-29T20:18:56 & 268.88 & 66.61 & 269.35 & 66.89\\
5020054.2 & IRC00 b;N & 2006-04-30T07:52:33 & 268.88 & 66.61 & 269.34 & 66.89\\
5020054.3 & IRC00 b;N & 2006-04-30T09:31:47 & 268.88 & 66.61 & 269.34 & 66.89\\
5020055.1 & IRC00 b;N & 2006-05-06T00:52:01 & 268.88 & 66.61 & 269.28 & 66.91\\
5020055.2\footnotemark[$*$] & IRC00 b;N & 2006-05-06T05:49:13 & 268.88 & 66.61 & 269.27 & 66.91\\
      \hline
\multicolumn{4}{@{}l@{}}{\hbox to 0pt{\parbox{180mm}{\footnotesize
 \footnotemark[$*$] MIR-L Data in this observation were not used.
}\hss}}
    \end{tabular}
  \end{center}
\end{table*}

\begin{table*}
  \caption{Summary of the survey}\label{tab:summary}
  \begin{center}
    \begin{tabular}{lllllll}
      \hline
      band & area         & $N_{F}$\footnotemark[$*$] & $T_{integ}$
     \footnotemark[$\dagger$] & $N_{S}$\footnotemark[$\ddagger$]&
     5$\sigma$ limit\footnotemark[$\S$]& 50\% limit\footnotemark[$\|$]\\
           & (arcmin$^2$) &          &       (sec)&         &  ($\mu$Jy) &  ($\mu$Jy)\\
     N3 & 86.0  & 102 & 4529 & 955 & 6.0  & 24.0\\
     S7 & 70.3  & 300 & 4908 & 298 & 31.5 & 47.5\\
     L15 & 77.3  & 270 & 4417 & 277& 71.2 & 88.1\\
     \hline
    \end{tabular}
  \end{center}
     \footnotemark[$*$] Number of frames co-added.\\
     \footnotemark[$\dagger$] Total integration time.\\
     \footnotemark[$\ddagger$] Number of source detected.\\
     \footnotemark[$\S$] Detection limit (sky noise limit).\\
     \footnotemark[$\|$] Completeness limit.\\
\end{table*}

\section{Data reduction}
 \label{sec-reduction}
The data were reduced by the standard IRC imaging pipeline (version
060801; see IRC Data User Manual, \cite{lorente-pipeline}).
The dark frame (version 060428), which was measured  just
before the PV phase observations began, was subtracted.
Flat fielding was performed based on the flat field frame (version 060626),
which was created from 6 pointed observations of a high background region
near the ecliptic plane with the AOT IRC02.
Distortion correction (version 060529) was performed based on the data 
obtained from the observations of globular clusters in the PV phase
\citep{ita-cluster}.
After the reduction of individual frames, we co-added
all the frames together.

In co-adding, we subtracted the self sky image from each individual frame
which was created by median filtering with 20$\times$20 pixel kernel, 
in order to remove the effect of stray light from the Earth shine.
Alignment of the frames was determined by the position
of the point sources in each frame in the case of the NIR and MIR-S channels.
In case of the MIR-L channel, the number of the point sources 
which had sufficient signal-to-noise ratio was not sufficient.
Therefore, in order to align the MIR-L frames, we used the alignment of the MIR-S frames, 
which were simultaneously observed with the MIR-L in a parallel field,
using the ``coaddLusingS'', an optional task in the IRC pipeline.
All the 10 pointing data  were co-added at once by the pipeline
in case of the NIR and MIR-S channel.
The task ``coaddLusingS'' which was used for the MIR-L data
does not support a rotation of the FOV and currently cannot handle multi-pointing
data.
Therefore, we co-added the MIR-L frames in each pointed observation
in order to improve the signal-to-noise ratio, and then co-added the frames
referring to positions of the point sources detected in each frame.

Any spurious events, such as cosmic ray events, 
moving debris, optical and electronic ghost images were removed
at the co-addition stage using a 3 sigma clipping technique.
We used the ``ccdclip'' option (in which the detector noise
performance were assumed) rather than the default option
``sigclip'' (in which the noise performance is estimated 
from the data itself). 
The average value among the frames, instead of the median value, for
each pixel was computed in order to improve the signal-to-noise ratio of the final image.

The number of frames which were successfully co-added were 102, 300 and 270 in
the N3, S7 and L15 bands, respectively (the L15 band data taken with the
target ID=5020055.2 was not used since the pipeline failed to co-add the
L15 data, probably because of large stray light from the Earth shine). One IRC pointing consists of long exposure frames together
with short exposure frames to obtain a wider dynamic range. However, we
co-added the long exposure frames only. 

Astrometry was based on a comparison of the positions 
of the point sources detected in the AKARI image and 
sources listed in the 2MASS point source catalog. 
For the N3 and S7 band images, a build-in function of the pipeline
successfully identified the AKARI sources with the sources in 2MASS catalog.
For the L15 band image, automatic matching failed 
due to the intrinsic red K - L15 color of the L15 sources.
We identified the 2MASS sources by eye for the L15 band images.
Cross identification with the Subaru/Suprime-Cam deep optical image 
 showed that the positional
accuracy was  better than 2 pixel.

\begin{figure*}
   \begin{center}
      \FigureFile(180mm,50mm){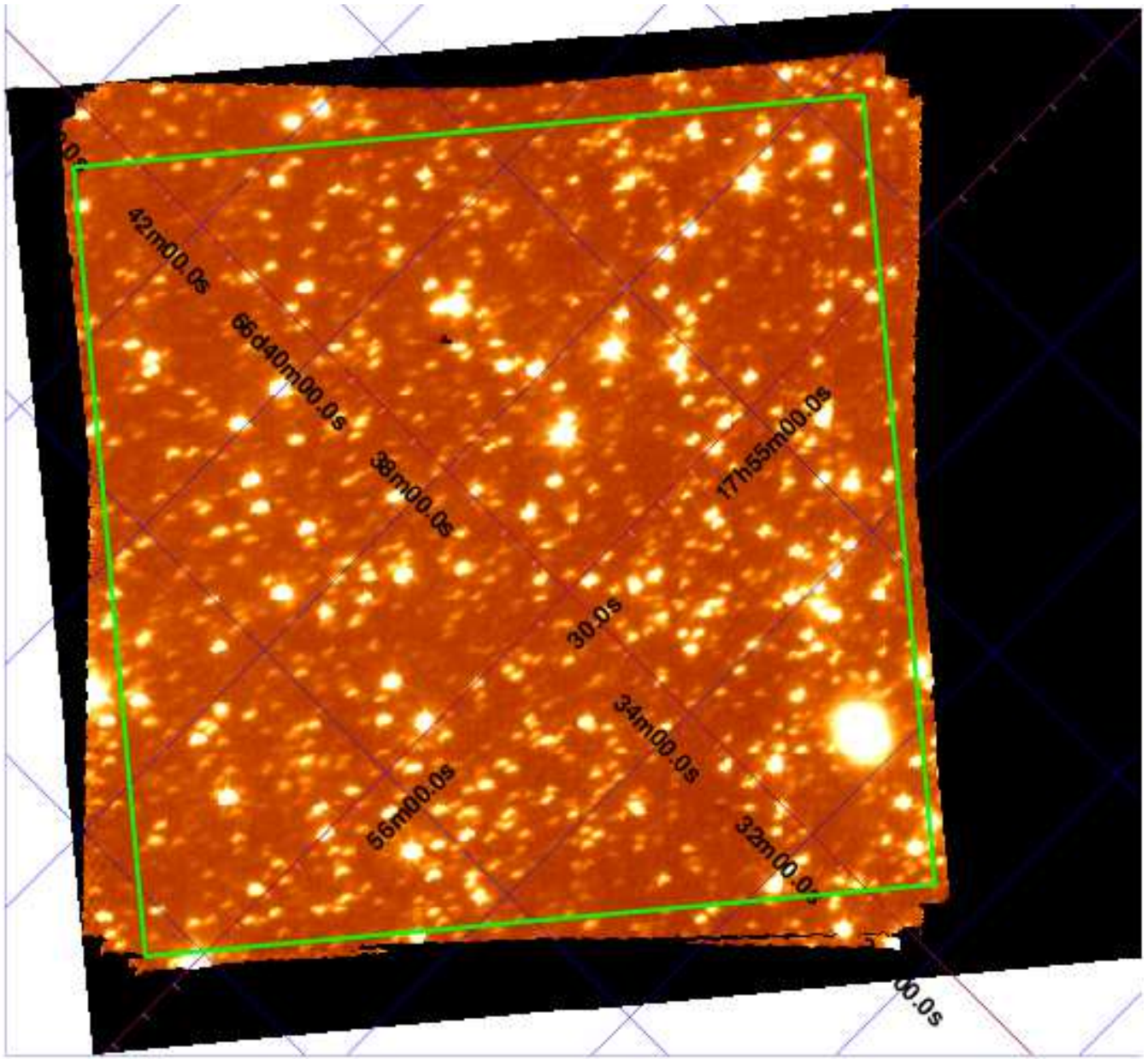}
   \end{center}
   \caption{
The final N3 co-added image is shown in equatorial coordinates (rotated
45 degrees counter-clockwise).
The green box indicates the region over which 
point source extraction was performed.
The size of region is 9.19 x 9.19 arcmin. 
}\label{fig:N3}
\end{figure*}

\begin{figure*}
   \begin{center}
      \FigureFile(180mm,50mm){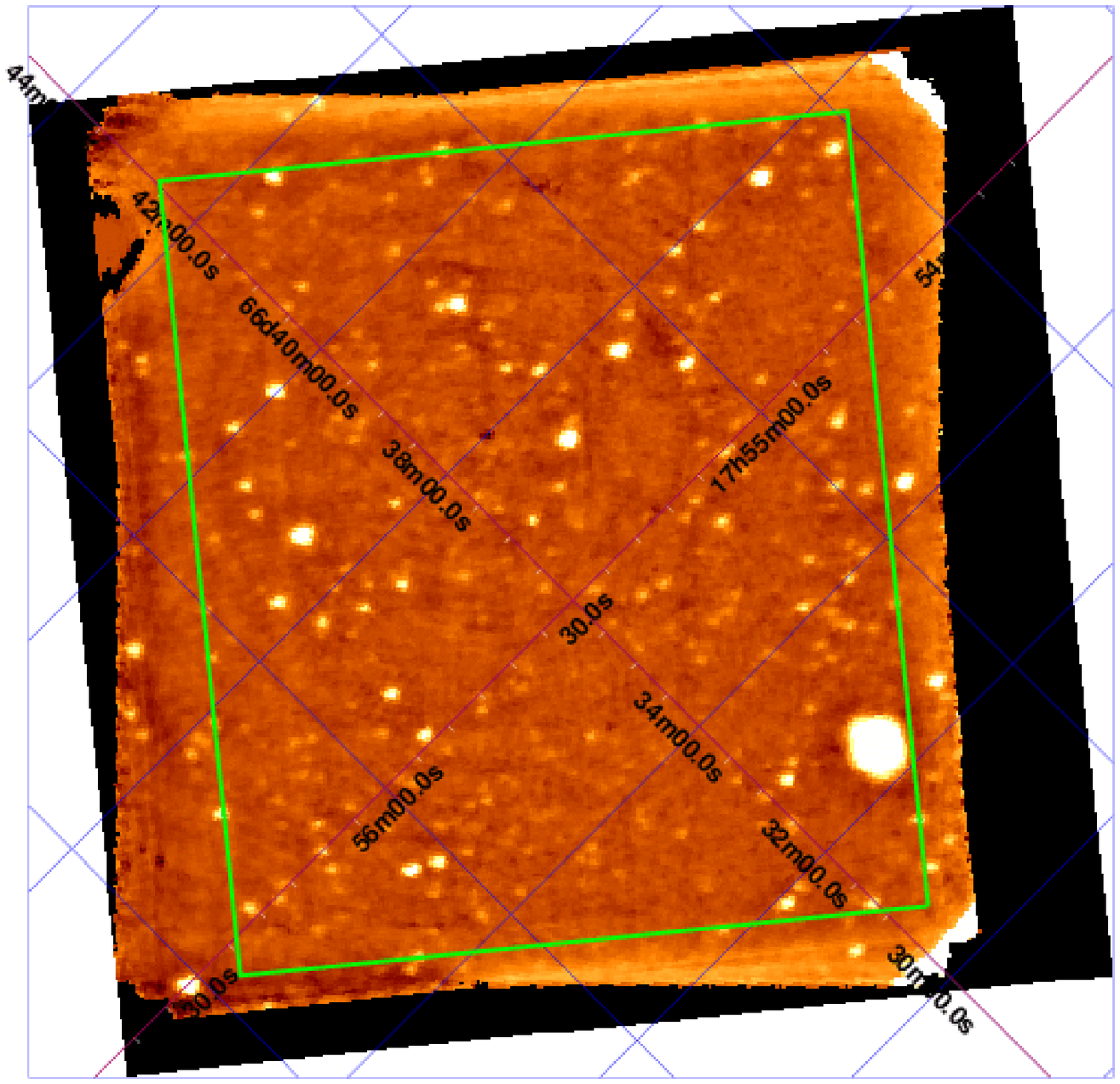}
   \end{center}
   \caption{
The final S7 co-added image is shown in equatorial coordinates (rotated
45 degrees counter-clockwise).
The green box indicates the region over which 
point source extraction was performed.
The size of region is 7.76 x 8.96 arcmin. 
}\label{fig:S7}
\end{figure*}

\begin{figure*}
   \begin{center}
      \FigureFile(180mm,50mm){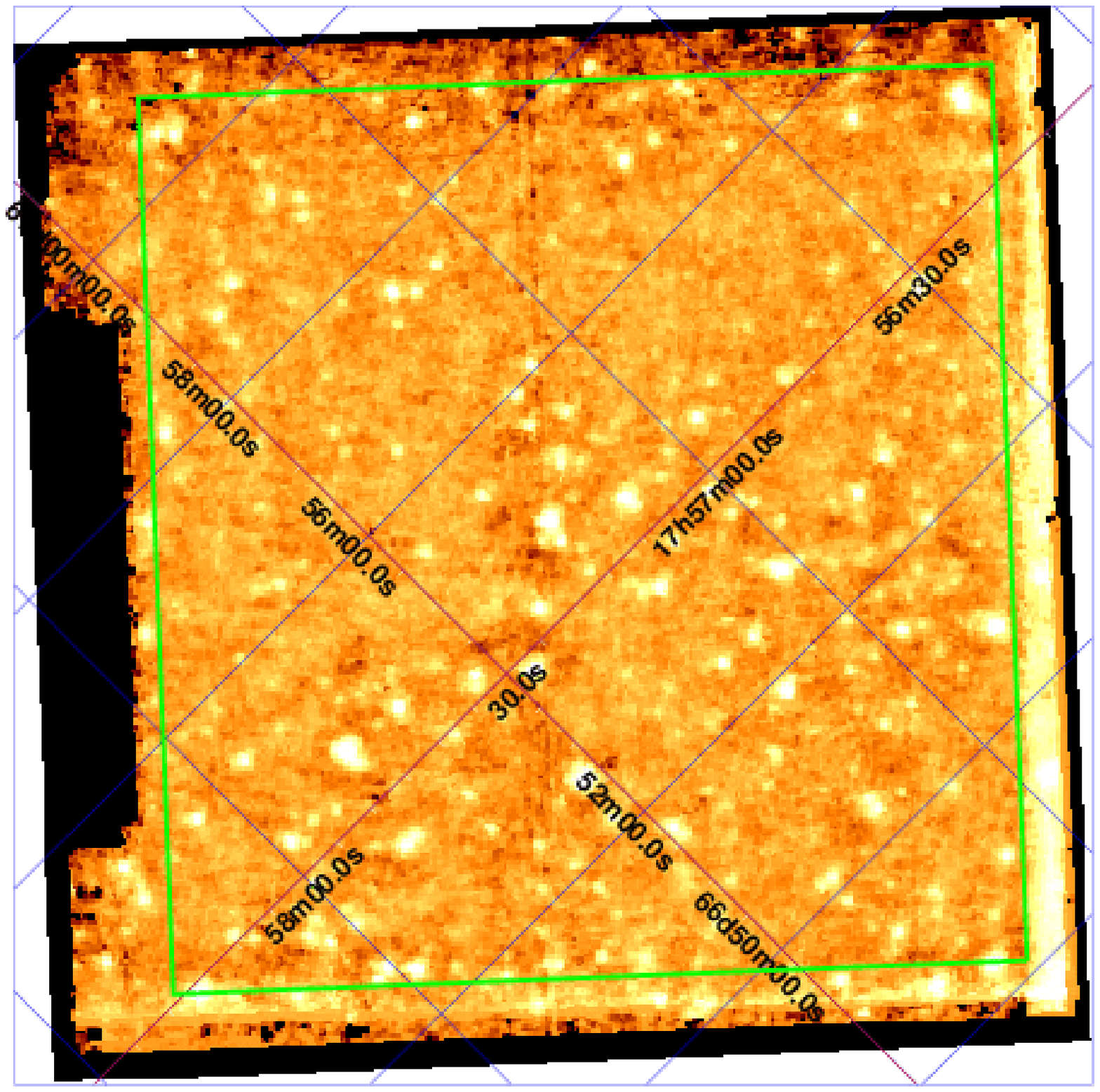}
   \end{center}
   \caption{
 The final L15 co-added image is shown in equatorial coordinates (rotated
45 degrees counter-clockwise).
The green box indicates the region over which 
point source extraction was performed.
 The size of region is 8.77 x 9.21 arcmin. 
 }\label{fig:L15}
\end{figure*}

Finally, we trimmed the area  close to the edge of each co-added
image in which the signal-to-noise ratio was poor.
The area of the final images are 85.95, 70.27 and 77.29 arcmin$^2$, for
the N3, S7, and L15 bands, respectively.
Figures~\ref{fig:N3}, \ref{fig:S7} and \ref{fig:L15} show 
the final images in the N3, S7, and L15 bands, respectively.

\begin{figure}
 \begin{center}
  \FigureFile(80mm,50mm){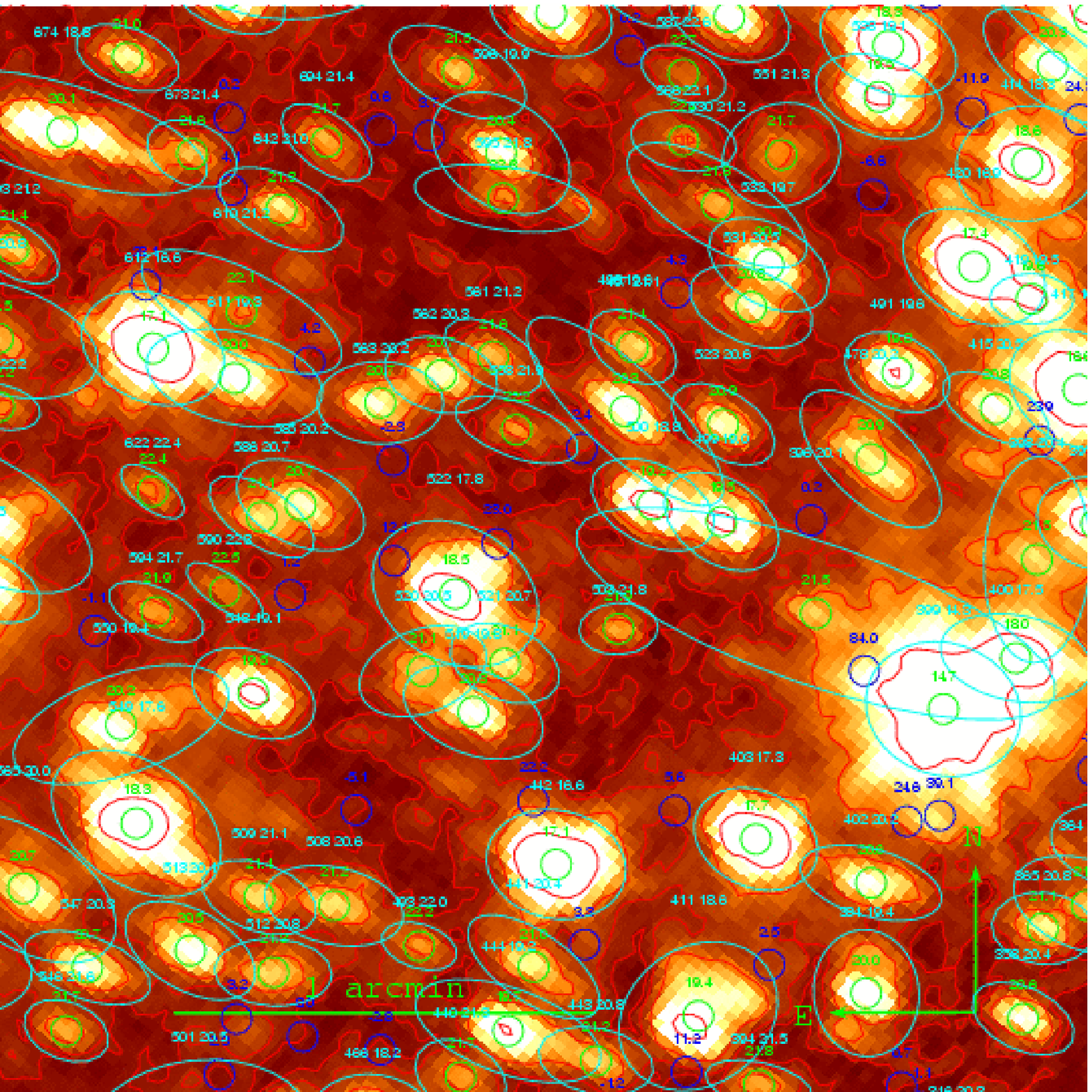}
 \end{center}
 \caption{
 Close up view of the AKARI/IRC N3 band image.
 Contours are 1, 3 and  10 sigma levels of the local background
 fluctuation.
 The green circles indicate the positions of detected sources.
 The MAGAUTO apertures (cyan) are superposed on each AKARI source.
 The blue circles indicate the apertures on blank sky positions 
 which were used for the calculation of the detection limit.
 }\label{fig:N3-apertures}
\end{figure}

\begin{figure}
 \begin{center}
  \FigureFile(80mm,50mm){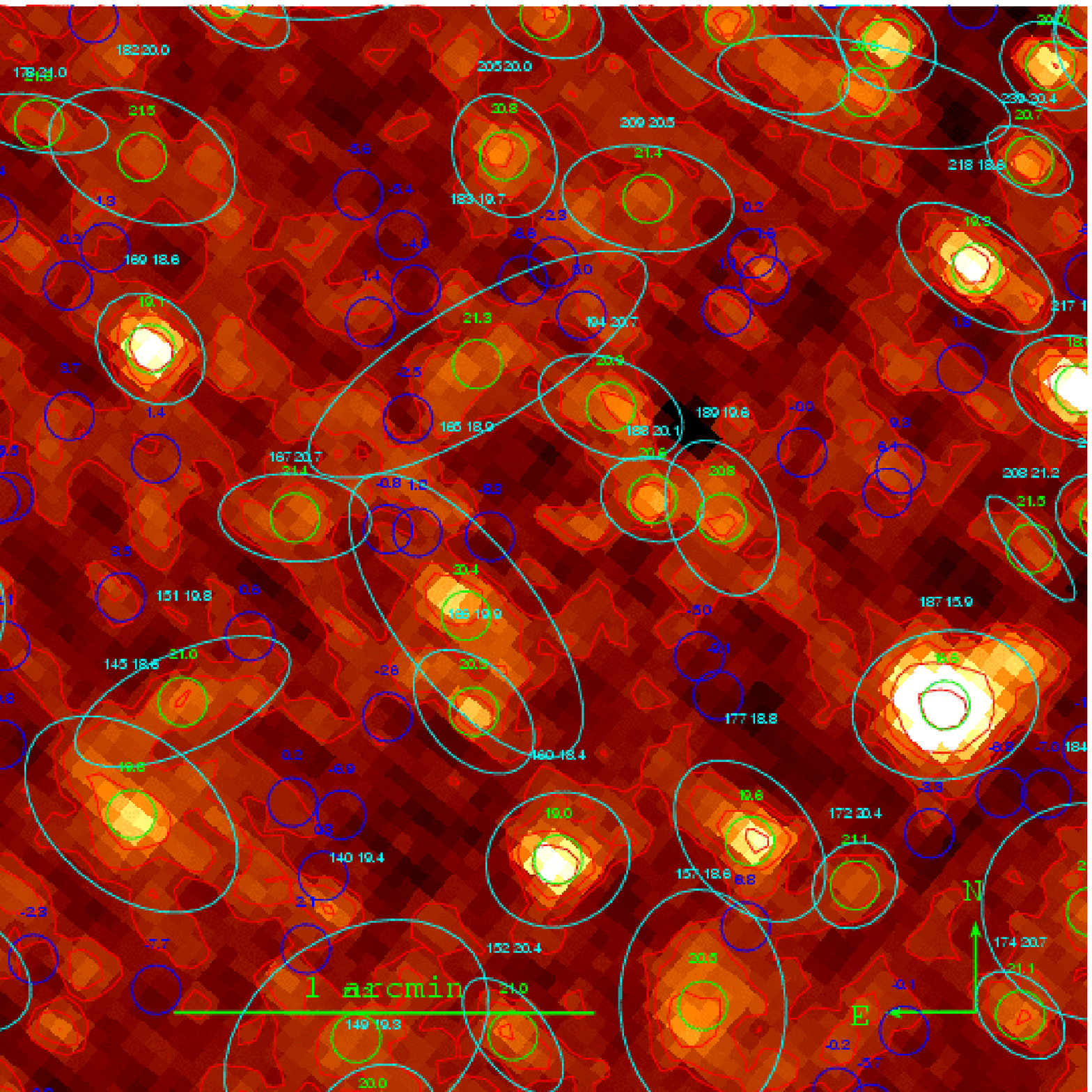}
 \end{center}
 \caption{
 Close up view of the AKARI/IRC S7 band image.
 Contours are 1, 3 and  10 sigma levels of the local background fluctuation.
 The green circles indicate the positions of detected sources.
 The MAGAUTO apertures (cyan) are superposed on each AKARI source.
 The blue circles indicate the apertures on blank sky positions 
 which were used for the calculation of the detection limit.
 Some of peaks which have not detected as a source by our simple 
 detection criterion have counter parts in the deep optical image.
 Further optimization of detection criterion
 may improve the detection limit.
 }\label{fig:S7-apertures}
\end{figure}

\begin{figure}
 \begin{center}
  \FigureFile(80mm,50mm){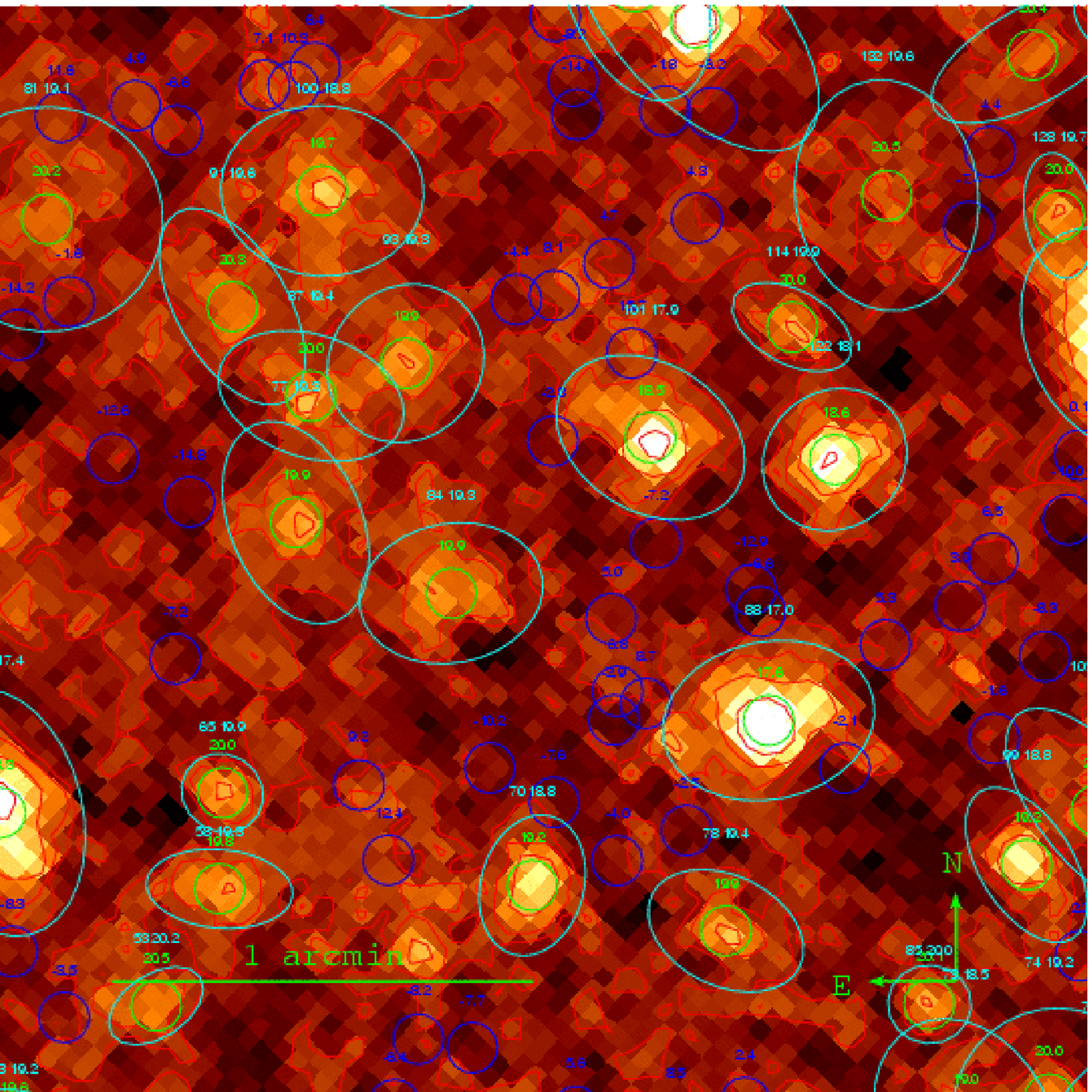}
 \end{center}
 \caption{
 Close up view of the AKARI/IRC L15 band image.
 Contours are 1, 3 and  10 sigma levels of the local background
 fluctuation.
 The green circles indicate the positions of detected sources.
 The MAGAUTO apertures (cyan) are superposed on each AKARI source.
 The blue circles indicate the apertures on blank sky positions 
 which were used for the calculation of the detection limit.
 }\label{fig:L15-apertures}
\end{figure}

 \section{Source extraction and photometry} \label{sec-catalog}
 Sextractor \citep{bertin-sextractor} was used for source extraction.
 The extraction criterion  was a connection of five pixels 
having more than a 1.6532 sigma significance above local sky fluctuation.
 If the flux is equally distributed over these five pixels,
 it corresponds to a 5 sigma detection. 
 The actual PSF has a weak concentration at its center, and the above 
criterion gives a slightly higher significance than a 5 sigma detection for
a point source.
 
The flux of each extracted source was evaluated by aperture photometry
calculated in Sextractor.
Elliptical aperture photometry with variable size (Sextractor's MAGAUTO)
was used for the further analysis. Parameters of ``Kron factor'' and ``minimum
radius'' were set to the default values of 2.5 and 3.5, respectively.
The actual apertures used for each detected source are overlaid 
on the AKARI/IRC images in the figures~\ref{fig:N3-apertures}, \ref{fig:S7-apertures} 
and~\ref{fig:L15-apertures}.
Magnitude zero point was derived by observations of standard
stars, and is described in \citet{lorente-pipeline}.

\begin{figure*}
   \begin{center}
      \FigureFile(50mm,50mm){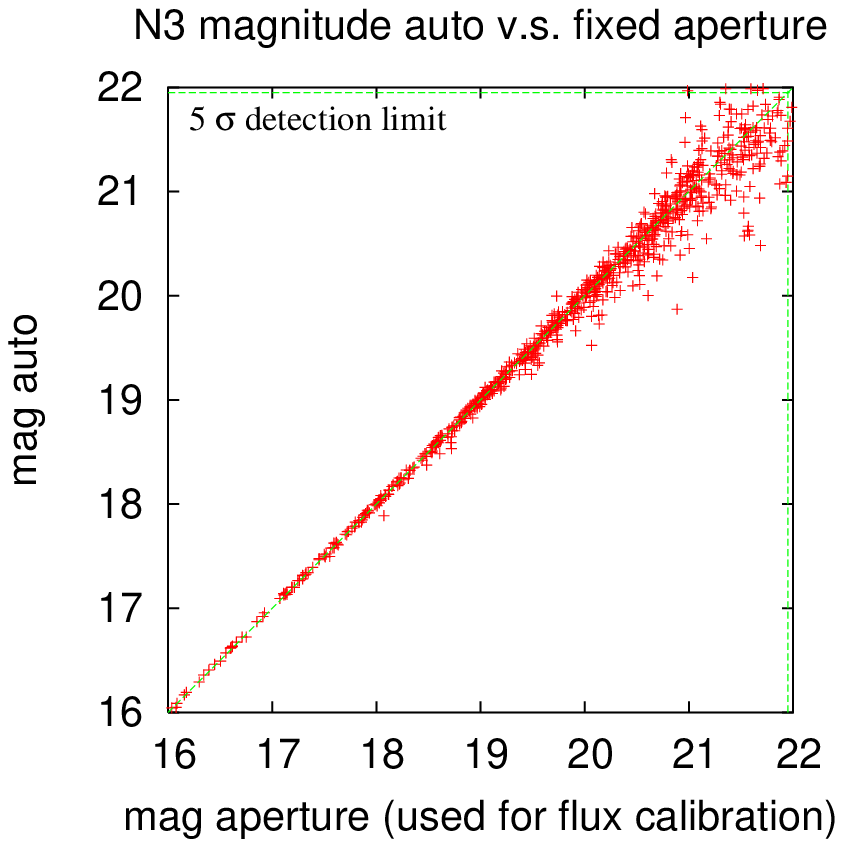}
      \FigureFile(50mm,50mm){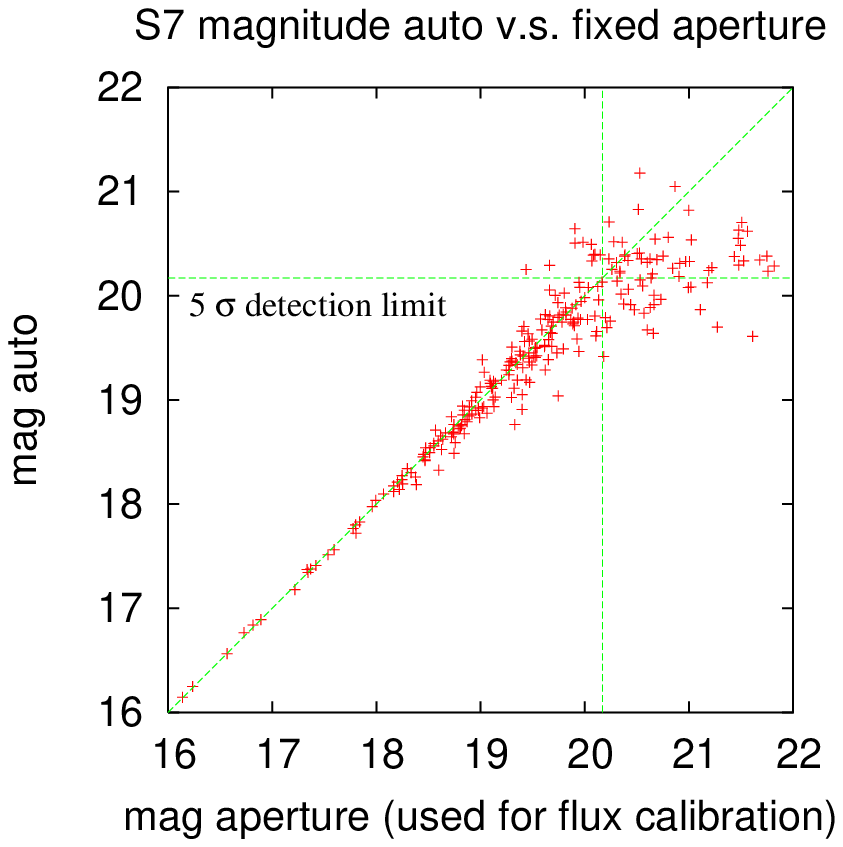}
      \FigureFile(50mm,50mm){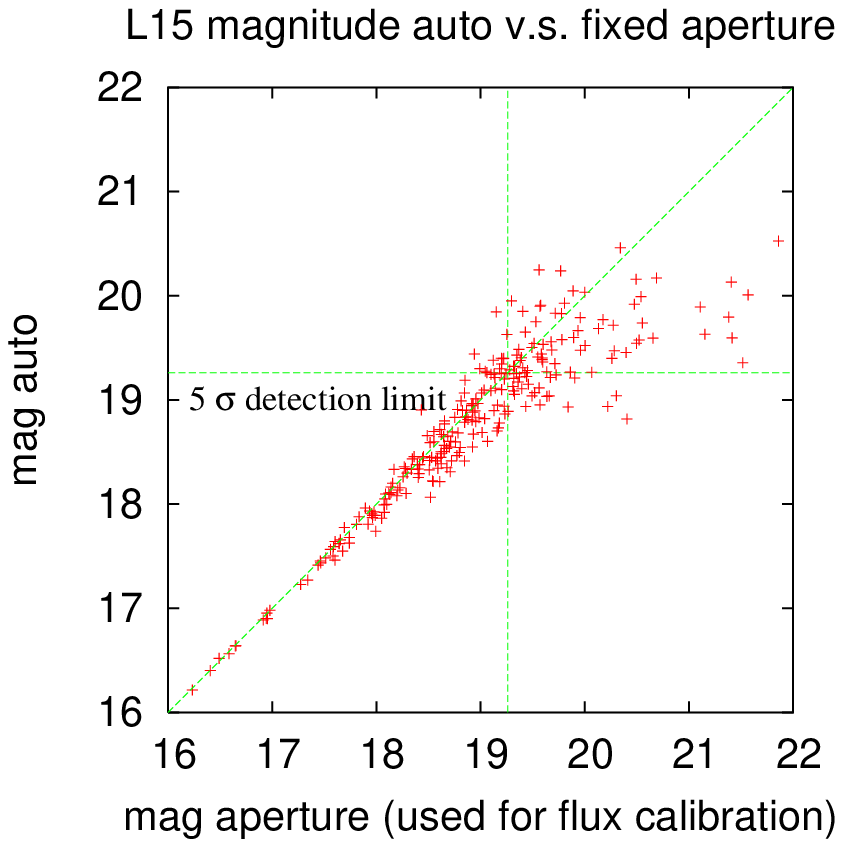}
   \end{center}
   \caption{
 Magnitudes of individual stars given by MAGAUTO are compared 
 with that by MAGAPER with the same aperture radius as the observations
 of the standard stars.
 No significant systematic difference between MAGAUTO and  MAGAPER is seen 
 above the 5 $\sigma$  detection limit, although there is large
 systematic difference at fainter level.
}
 \label{fig:magauto-magaper}
\end{figure*}

Simple aperture photometry (Sextractor's MAGAPER) was also calculated 
in order to estimate the amount of aperture correction.
The aperture radius was set to the same as used in the standard star
observations, 10 and 7.5 pixels in the NIR and MIR-S/L images, respectively. 
Figure \ref{fig:magauto-magaper} shows the comparison between
the flux estimations by MAGAUTO and MAGAPER for each source.
We did not apply any aperture correction for further analysis
because the results using MAGAUTO and MAGAPER were reasonably consistent.

\begin{figure*}
   \begin{center}
    \FigureFile(80mm,50mm){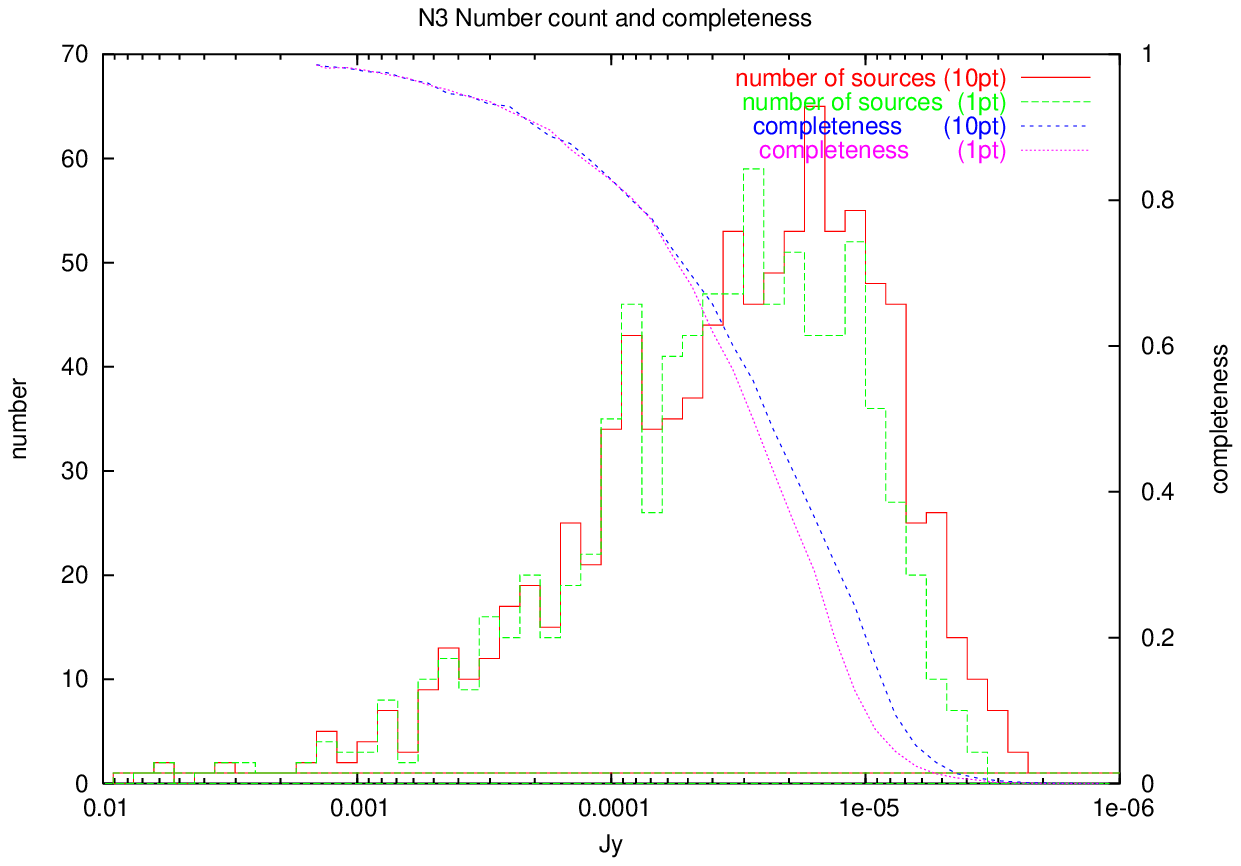}
    \FigureFile(80mm,50mm){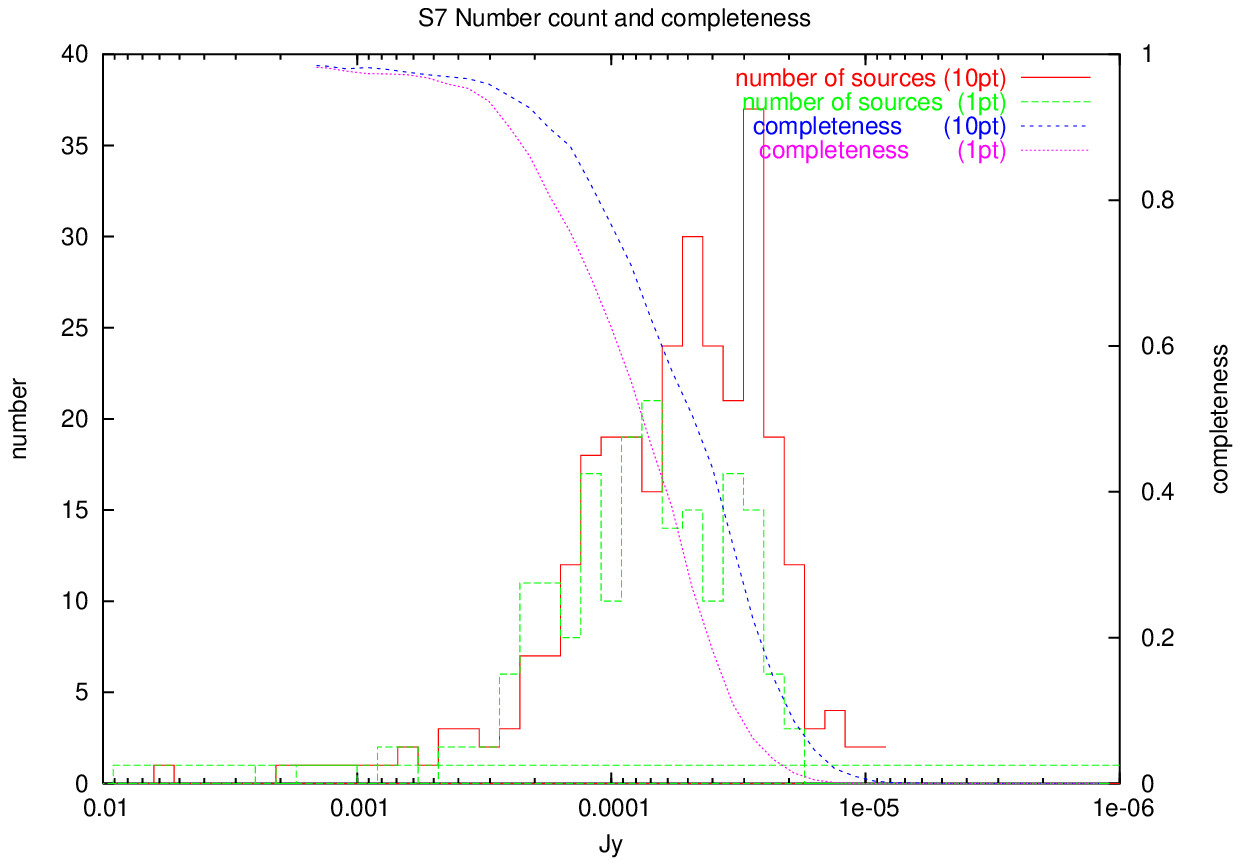}
    \FigureFile(80mm,50mm){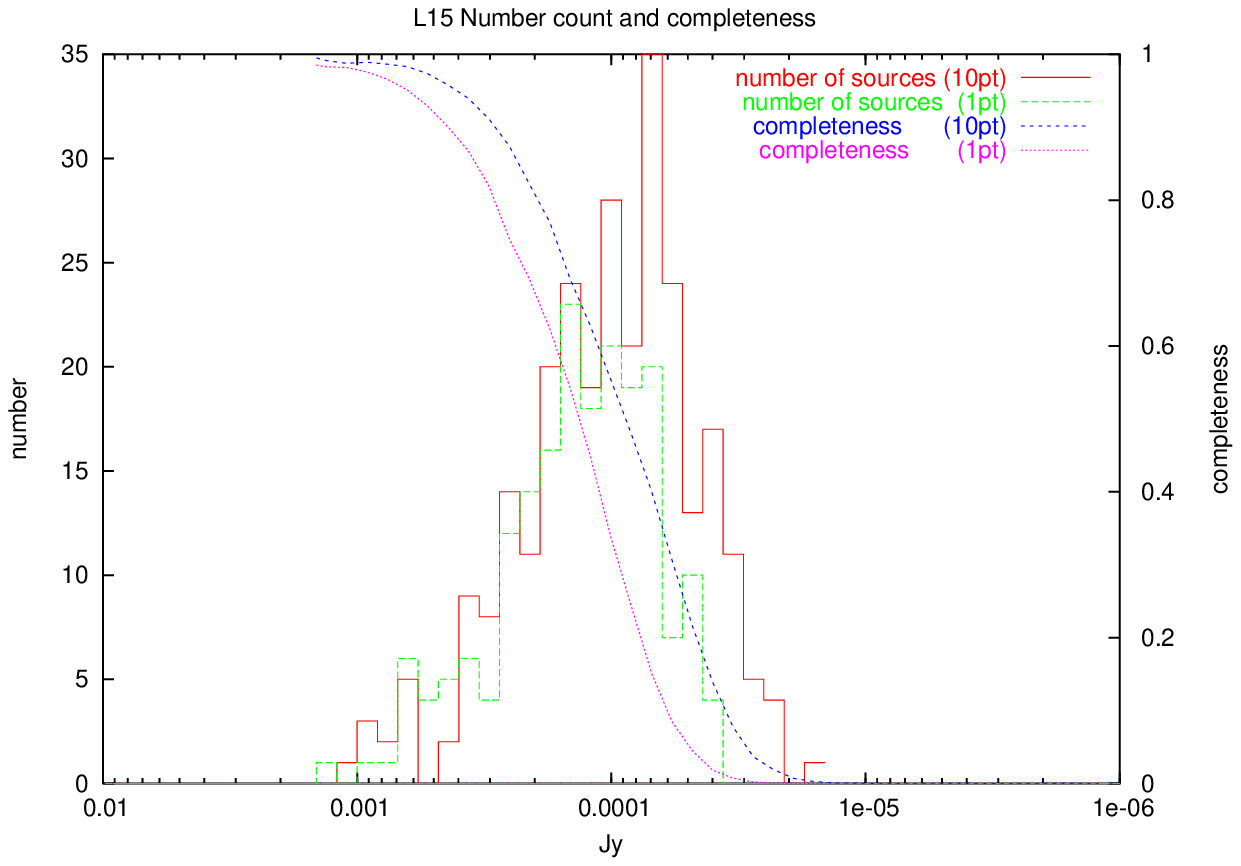}
   \end{center}
   \caption{
Differential source counts and completeness estimation
is shown for the N3, S7, and L15 bands, respectively.
The results from both a single pointed and ten pointed observations
are plotted.
}
\label{fig:nc-comp}
 
\end{figure*}

\begin{table*}
  \caption{Source counts and completeness}\label{tab:nc-comp}
  \begin{center}
     \begin{tabular}{rrrllrllrll}
      \hline
         &           & \multicolumn{3}{c}{N3}&\multicolumn{3}{c}{S7}&\multicolumn{3}{c}{L15} \\
mag$_{AB}$  & log(Jy) &N\footnotemark[$*$] &$\overline{C}$\footnotemark[$\dagger$]& $\sigma(C)$\footnotemark[$\ddagger$]& N &$\overline{C}$ & $\sigma(C)$ & N  &$\overline{C}$  & $\sigma(C)$ \\
$<$15.8 & $> -2.76$ &     10 &  & &    3 &  & &     0 &  & \\
15.8--16.0 & $-2.80$&     2 & 0.986 & 0.007 &     1 & 0.985 & 0.006 &     0 & 0.995 & 0.002 \\
16.0--16.2 & $-2.88$&     5 & 0.983 & 0.005 &     1 & 0.984 & 0.006 &     0 & 0.992 & 0.005 \\
16.2--16.4 & $-2.96$&     2 & 0.981 & 0.005 &     1 & 0.980 & 0.005 &     1 & 0.988 & 0.005 \\
16.4--16.6 & $-3.04$&     4 & 0.976 & 0.006 &     1 & 0.982 & 0.005 &     3 & 0.989 & 0.006 \\
16.6--16.8 & $-3.12$&     7 & 0.975 & 0.008 &     1 & 0.979 & 0.008 &     2 & 0.987 & 0.006 \\
16.8--17.0 & $-3.20$&     3 & 0.966 & 0.009 &     2 & 0.975 & 0.007 &     5 & 0.983 & 0.005 \\
17.0--17.2 & $-3.28$&     9 & 0.961 & 0.009 &     1 & 0.972 & 0.009 &     0 & 0.973 & 0.005 \\
17.2--17.4 & $-3.36$&    13 & 0.946 & 0.009 &     3 & 0.969 & 0.007 &     2 & 0.958 & 0.006 \\
17.4--17.6 & $-3.44$&    10 & 0.943 & 0.010 &     3 & 0.966 & 0.008 &     9 & 0.940 & 0.009 \\
17.6--17.8 & $-3.52$&    12 & 0.932 & 0.010 &     2 & 0.959 & 0.009 &     8 & 0.911 & 0.014 \\
17.8--18.0 & $-3.60$&    17 & 0.929 & 0.010 &     3 & 0.943 & 0.011 &    14 & 0.874 & 0.018 \\
18.0--18.2 & $-3.68$&    19 & 0.907 & 0.015 &     7 & 0.927 & 0.012 &    11 & 0.822 & 0.020 \\
18.2--18.4 & $-3.76$&    15 & 0.888 & 0.015 &     7 & 0.898 & 0.015 &    20 & 0.770 & 0.023 \\
18.4--18.6 & $-3.84$&    25 & 0.878 & 0.016 &    12 & 0.874 & 0.016 &    24 & 0.692 & 0.016 \\
18.6--18.8 & $-3.92$&    21 & 0.854 & 0.019 &    18 & 0.822 & 0.013 &    19 & 0.627 & 0.022 \\
18.8--19.0 & $-4.00$&    34 & 0.829 & 0.019 &    19 & 0.767 & 0.022 &    28 & 0.554 & 0.023 \\
19.0--19.2 & $-4.08$&    43 & 0.801 & 0.021 &    19 & 0.711 & 0.023 &    21 & 0.480 & 0.026 \\
19.2--19.4 & $-4.16$&    34 & 0.775 & 0.021 &    16 & 0.635 & 0.026 &    35 & 0.401 & 0.016 \\
19.4--19.6 & $-4.24$&    35 & 0.734 & 0.024 &    24 & 0.566 & 0.028 &    24 & 0.307 & 0.025 \\
19.6--19.8 & $-4.32$&    37 & 0.696 & 0.021 &    30 & 0.505 & 0.029 &    13 & 0.217 & 0.022 \\
19.8--20.0 & $-4.40$&    44 & 0.657 & 0.016 &    24 & 0.432 & 0.021 &    17 & 0.139 & 0.014 \\
20.0--20.2 & $-4.48$&    53 & 0.601 & 0.021 &    21 & 0.328 & 0.022 &    11 & 0.078 & 0.018 \\
20.2--20.4 & $-4.56$&    46 & 0.552 & 0.030 &    37 & 0.225 & 0.018 &     5 & 0.037 & 0.010 \\
20.4--20.6 & $-4.64$&    49 & 0.485 & 0.027 &    19 & 0.145 & 0.020 &     4 & 0.019 & 0.006 \\
20.6--20.8 & $-4.72$&    53 & 0.426 & 0.028 &    12 & 0.086 & 0.014 &     0 & 0.007 & 0.005 \\
20.8--21.0 & $-4.80$&    65 & 0.365 & 0.031 &     3 & 0.046 & 0.010 &     1 & 0.003 & 0.003 \\
21.0--21.2 & $-4.88$&    53 & 0.303 & 0.024 &     4 & 0.020 & 0.006 &       & 0.001 & 0.001 \\
21.2--21.4 & $-4.96$&    55 & 0.243 & 0.019 &     2 & 0.010 & 0.005 &       & 0.000 & 0.001 \\
21.4--21.6 & $-5.04$&    48 & 0.165 & 0.020 &     2 & 0.003 & 0.002 &       & 0.000 & 0.000 \\
21.6--21.8 & $-5.12$&    46 & 0.094 & 0.012 &       & 0.001 & 0.002 &       & 0.000 & 0.000 \\
21.8--22.0 & $-5.20$&    25 & 0.052 & 0.014 &       & 0.000 & 0.001 &       & 0.000 & 0.000 \\
22.0--22.2 & $-5.28$&    26 & 0.029 & 0.009 &       & 0.000 & 0.001 &       & 0.000 & 0.000 \\
22.2--22.4 & $-5.36$&    14 & 0.014 & 0.006 &       & 0.000 & 0.000 &       & 0.000 & 0.001 \\
22.4--22.6 & $-5.44$&    10 & 0.008 & 0.004 &       & 0.000 & 0.001 &       & 0.000 & 0.001 \\
22.6--22.8 & $-5.52$&     7 & 0.004 & 0.003 &       & 0.000 & 0.000 &       & 0.000 & 0.001 \\
22.8--23.0 & $-5.60$&     3 & 0.002 & 0.002 &       & 0.000 & 0.000 &       & 0.000 & 0.000 \\
23.0--23.2 & $-5.68$&     1 & 0.001 & 0.001 &       & 0.000 & 0.000 &       & 0.000 & 0.000 \\
     \hline
    \end{tabular}
 \end{center}
\footnotemark[$*$]  Number of source detected in each magnitude bin.\\
\footnotemark[$\dagger$] Average of completeness estimation in 25 Monte-Carlo simulations.\\
\footnotemark[$\ddagger$] Standard deviation of completeness estimation
 for the 25  Monte-Carlo simulations.\\
\end{table*}

\section{Estimation of sensitivity and completeness}
\label{sec-sensitivity}

\subsection{detection limit (sky noise limit)}
The sensitivity of the point source extraction
was estimated by measuring the fluctuation of the photometry 
on blank sky at random position (sky noise).
The positions which were close to a source were not used for the measurement
in order to avoid source contamination effects.

We made simple aperture photometry (a no weighting, no PSF fitting
technique was applied, and no centering was performed) 
at each random position using the IRAF/PHOT package.
The size of aperture radius was set to 1.5 pixel for all three bands.
The aperture had slightly wider area (7.0 pixels) 
than the minimum number of connected pixels 
in the extraction criterion (5 pixels). 
a slightly worse estimation of the sensitivity resulted from this choice of
aperture size.
The actual apertures which were used for above measurement are shown
in the figures~\ref{fig:N3-apertures}, \ref{fig:S7-apertures}
and~\ref{fig:L15-apertures}.

In order to evaluate the fluctuation of photometry,
the histogram of the photometry was fit with a Gaussian function.
We define the one sigma level of the fluctuation 
as the standard deviation of the Gaussian.
Finally, taking into account the aperture correction,
we calculate a 5 sigma detection limit for a point source.

The 5 sigma detection limits are estimated to be  11.2, 71.3 and 201 $\mu$Jy 
for the co-added images obtained in a single pointed observation, 
and 6.0, 31.5 and 71.2 $\mu$Jy for that in the ten pointed observations,
in the N3, S7, L15 bands, respectively.
The results of the detection limit estimation are
 summarized in Table~\ref{tab:limits}. 

The improvements from one to ten pointings
were a factor of 1.87, 2.27 and 2.82 in the N3, S7, and L15 bands, 
respectively.
These factors in the N3 and S7 bands are slightly worse than what is expected
from the increase of number of observations (i.e., square root of ten).

\begin{table*}
  \caption{Detection limit (sky noise limit) and Completeness limit in $\mu$Jy}\label{tab:limits}
  \begin{center}
\begin{tabular}{|c|c|c|c|c|c|c|c|c|c|c|c|c|} \hline 
filter & \multicolumn{4}{c|}{N3} & \multicolumn{4}{c|}{S7} & \multicolumn{4}{c|}{L15}  \\ \hline
limit & \multicolumn{3}{c|}{completeness }& detection &
 \multicolumn{3}{c|}{completeness }& detection  &
 \multicolumn{3}{c|}{completeness }& detection  \\ \hline
criterion &30\% &50\% &80\% & 5$\sigma$ &30\% &50\% &80\% & 5$\sigma$ &30\% &50\% &80\% & 5$\sigma$  \\ \hline
1pt& 16.3 & 27.7 & 81.9 & 11.2 & 50.8 & 75.5 & 171.4 & 71.3 & 93.6 & 134.2 & 289.6 & 200.9 \\  \hline
10pt& 13.1 & 24.0 & 83.0 & 6.0 & 31.7 & 47.5 & 112.5 & 31.5 & 57.0 & 88.1 & 195.2 & 71.2 \\  \hline
1pt/10pt& 1.24 & 1.15 & 0.99 & 1.87 & 1.60 & 1.59 & 1.52 & 2.27 & 1.64 & 1.52 & 1.48 & 2.82 \\  \hline
\end{tabular}
  \end{center}
\end{table*}

\subsection{completeness correction}
The completeness of the source extraction was estimated via Monte Carlo
simulations.
We added artificial sources from a truth catalog into the final co-added
images,
and then extracted the sources again using the same extraction parameters.
We, then, compared the position and magnitude of input sources with the
extracted sources.
If there was a extracted source within 2 pixels around an input source 
and if the difference in magnitude  between the input source
and the extracted source was less than 0.5 magnitude,
we regarded the event as a successful extraction of an artificial truth
source, and counted it.

We define the completeness of the source extraction
as the number of successfully extracted truth sources
divided by the total number of artificial input sources
that were originally added into the original image.

The calculation was performed by dividing our flux range into magnitude bins
of size 0.2 magnitudes (i.e. uniform bin size in log Jy) and populating each
bin with 20 artificial sources. The magnitude distribution of the sources
was normalized to a flat Euclidian
universe in each bin. These 20 sources were then added in random positions
in the image, assuming a circular PSF. In order to avoid self-confusion
effects among the artificial sources, we did not insert a source if the
minimum distance to another artificial source 
was less than 20 pixels. Note that the number of random sources that
can be input at any one time is constrained by the array size of the IRC
(NIR: 412$\times$512 pixels, MIR-S/L:256$\times$256 pixels).

In order to improve the statistical significance
of the Monte Carlo simulations, we performed 20 statistically
independent calculations for each magnitude bin, creating an effective total
of 400 artificial sources. The simulation was then subsequently carried out
for each magnitude bin in turn.

In order to estimate the error in the calculation of the
completeness, we then performed 25 statistically independent sets
of the above simulations.
The results are shown in the figure \ref{fig:nc-comp}, and
the completeness limits are summarized in table~\ref{tab:limits}.

\section{discussion}
\label{sec-discussion}

\subsection{improvement of sensitivity from 1pt to 10pt}
Simple statistics expect a factor of $\sqrt{10/1}$ improvement
in sensitivity for 10 pointed observations compared 
to that for a single pointed observation. 
However, table~\ref{tab:limits} shows that
the sensitivity did not improve as well as expected.

\begin{table*}
 \caption{source number density in the unit of source per beam}\label{tab:beam-source}
 \begin{center}
  \begin{tabular}{ccccc} \hline 
   filter  & \multicolumn{2}{c}{the number of source per
   beam\footnotemark[$*$]} & \multicolumn{2}{c}{FWHM of PSF} \\ \hline
           & 1 pt  & 10pt       & in pixel & in arcsec\\ \hline
   N3      & 1/36.6  & 1/31.8   & 2.9 & 4.2 \\  \hline
   S7      & 1/76.7  & 1/56.3 & 2.2 & 5.2\\  \hline
   L15     & 1/81.7  & 1/61.0 & 2.3 & 5.5\\  \hline
  \end{tabular}
 \end{center}
 \footnotemark[$*$] The radius of the beam is set to FWHM/2.35 of the PSF.
\end{table*}

One of the effects which degrades the sensitivity at the completeness limit is 
source confusion. 
In order to estimate the effect of source confusion,
we calculate the source number density in the unit 
of source per beam(Table~\ref{tab:beam-source}).
In the case of the N3 band, the density is nearly equal to 1/30,
the classical limit of source confusion (\cite{condon74}; \cite{hogg01}). 
Clearly, the observations are limited by the source confusion,
and this is the reason why there is almost no improvement 
in the sensitivity from 1 to 10 pointed observations.
In the case of the S7 and L15 bands, 
the density is still lower than the classical confusion limit, 
and there is still some room for improvement of the sensitivity
by further pointed observations.

The detection limit in the L15 band shows a factor of 3 
improvement which is almost as expected from simple statistics,
while the detection limits in the N3 and S7 show somewhat
smaller improvements than expected.
The detection limit is estimated by the fluctuation of the local
background where there are no sources detected.
So, a smaller improvement in the detection limit means 
that the signal to noise ratio of the sky level is not improved by
multiple observations, or alternatively that there are real fluctuations in the sky.

There are two possible causes which suppress the improvement
in the S/N ratio of the sky level.
(1) Incomplete sky subtraction in the co-addition stage in the pipeline
  creates additional noise in the final image.
  There are two problems which may affect the sky subtraction.
  The first is stray light from the Earth shine, 
  which was not uniform over the FOV.
  The other is scattered light from the edge of the detector.
  Both create non-uniformity in the background and may prevent
  improvement by co-adding the frames. In order to eliminate these effects,
  self sky subtraction was performed, i.e. we subtracted the median
  value of surrounding 20$\times$20 pixels from the value of each pixel,
  but there may still be some residual.
  The modeling of the stray light from the Earth shine is still ongoing
  and any data taken in the northern sky during the months of May, 
  June and July, may suffer from the Earth shine problem, 
  and must be handled with great care.
  We have modeled the scattered light from the edge of the detector,
  and this model will be included in the next release of 
  the IRC imaging ``pipeline''.
(2) Incomplete dithering in position in multiple observations.
In order to eliminate systematic errors such as flat fielding error, 
redundant observations, in which observations are performed by independent
detector pixels, are necessary. 
For this purpose, dithering of the target position was performed.
The AOT IRC00 which we used in this observation  did not perform
any dithering during each single pointed observation. 
Dithering of the target position must be 
realized among several pointed observations. 
Since these observations were executed in the PV phase, 
dithering operations among pointed observations was not implemented, at
that time\footnote{ The specification of the absolute pointing accuracy
of the AKARI telescope is 30 arcsec. We therefore expect that 
dithering in position between multiple pointed observations is achieved
by scattering from positional error.
Actual absolute pointing accuracy, in repeated observation for a single
target position, is measured by this observation, and is turned out 
to be better than that of expected (the error is within FWHM of the PSF).
After this observations, we implemented dithering for multi-pointed
observation by shifting the target position for each pointing.}.
As described in section~\ref{sec-observation}, only a circular change of the
target position has been performed in the N3 and S7 band observations, 
and one directional shifting of target position has been
performed in the L15 band observations. 
Systematic errors in the co-addition of frames may
result from this incompleteness in the dithering.
We have since implemented dithering among later pointed observations, and 
the observations taken after September 2006 do not suffer problems 
in dithering.

There is some possibility that we have detected real sky fluctuations.
Fluctuation analysis of the background, however, is still underway, 
and further deeper observations (with a smaller PSF) 
shall be made in order to produce a clear answer to this issue. 
One thing we have to take care is that there may be some
sources which have significant signal to noise ratio,
but could not be extracted by our detection criterion (see figure \ref{fig:S7-apertures}).
Comparison between the deep optical image and the AKARI S7 image shows
that there are optical counterpart for some faint
(below the detection limit) AKARI sources. 
The detection criterion which is used in this work is 
very simple and there is much room for improvement.
Further optimization of the detection criterion 
may improve the detection limit.

\subsection{3 and 7 $\mu$m source counts}

\begin{figure*}
   \begin{center}
      \FigureFile(160mm,50mm){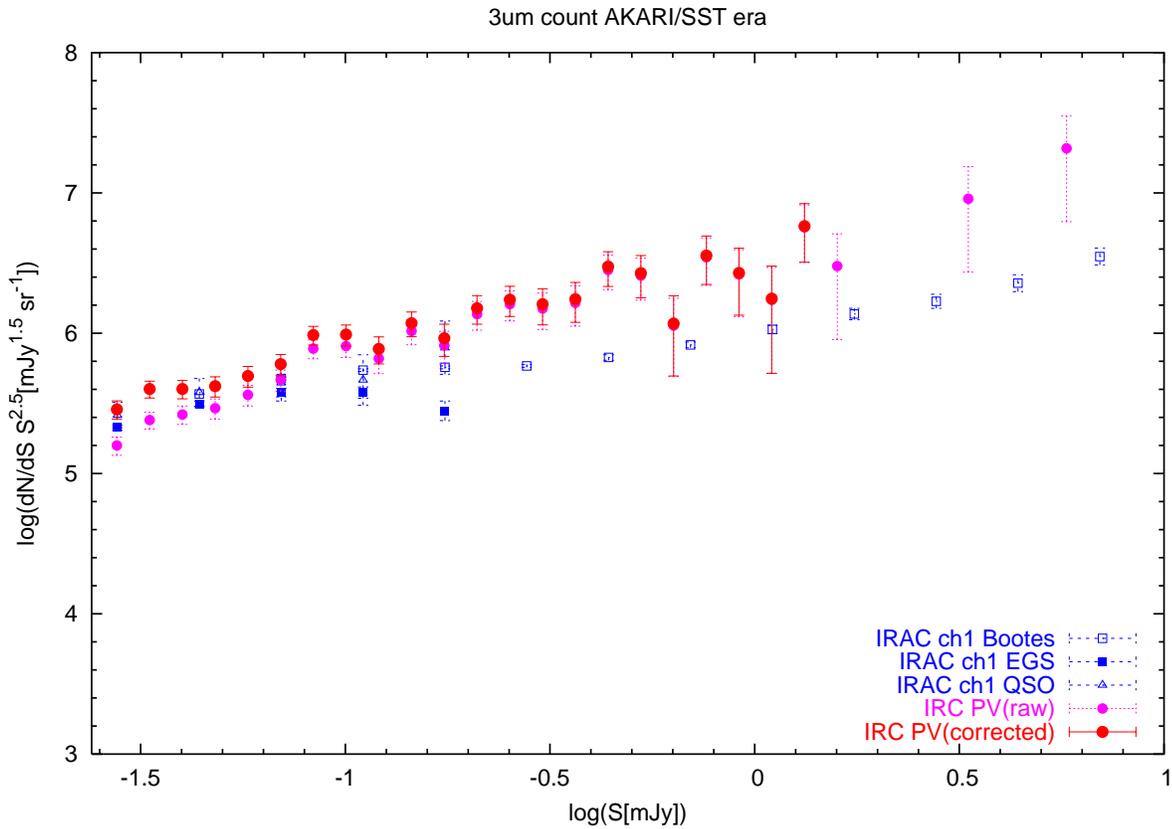}
   \end{center}
   \caption{
AKARI/IRC 3 $\mu$m differential source counts (both raw and
corrected for in-completeness) are plotted with 
the surveys carried out by Spitzer/IRAC results.
The source counts are normalized to a Euclidian  flat universe.
Only flux density level above 50\% completeness limit of our results
is shown.
IRAC channel 1 (3.6 $\mu$m) results (corrected for in-completeness) are
 from \citet{fazio-irac}.
}\label{fig:lognlogs_n3_wide}
\end{figure*}

\begin{figure*}
   \begin{center}
      \FigureFile(160mm,50mm){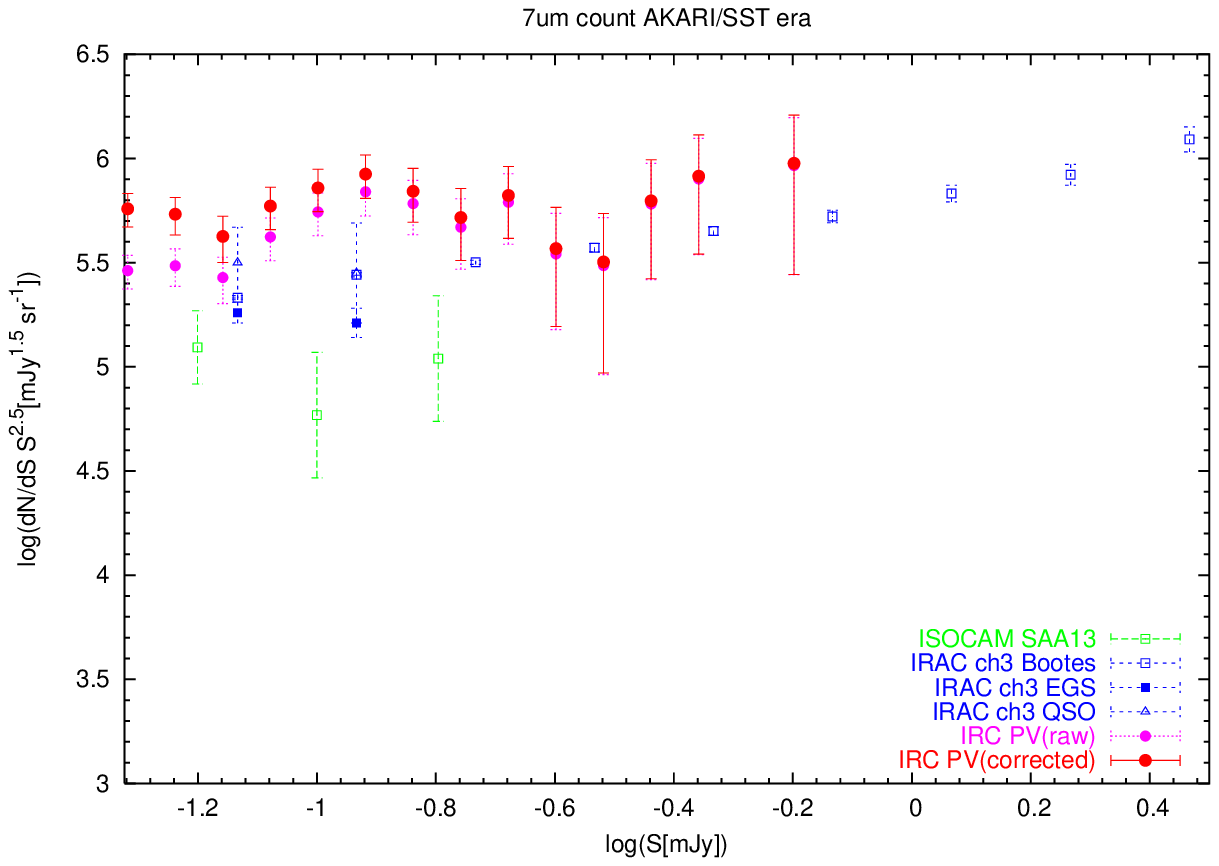}
   \end{center}
   \caption{
AKARI/IRC 7 $\mu$m differential source counts (both raw and
corrected for in-completeness) are plotted with 
the surveys carried out by  ISO/ISOCAM and Spitzer/IRAC results.
The source counts are normalized to a Euclidian  flat universe.
Only flux density level above 50\% completeness limit of our results
is shown.

IRAC channel 3 (5.6 $\mu$m) results (corrected for in-completeness) are from \citet{fazio-irac}.
ISOCAM LW2 (6.7 $\mu$m) results (raw data) are from \citet{sato03}.
}\label{fig:lognlogs_s7_wide}
\end{figure*}

Figures~\ref{fig:lognlogs_n3_wide} and~\ref{fig:lognlogs_s7_wide} 
show the 3 and 7  $\mu$m  number count results of the  AKARI/IRC,
ISO/ISOCAM and {\it Spitzer} IRAC surveys. 
The counts corrected for in-completeness are consistent with each other 
considering the cosmic variance.

The surveys by IRAC  reached deeper than 
those of the IRC because of the larger aperture size of telescope 
and smaller size of the PSF.
The IRC, however, has a larger FOV (4 times larger than that of IRAC) and
a larger number of filters (6 filters from 2 to 11 $\mu$m),
hence has a large merit in wide field and multi-band surveys.

\subsection{15 $\mu$m source counts}

\begin{figure*}
   \begin{center}
      \FigureFile(160mm,50mm){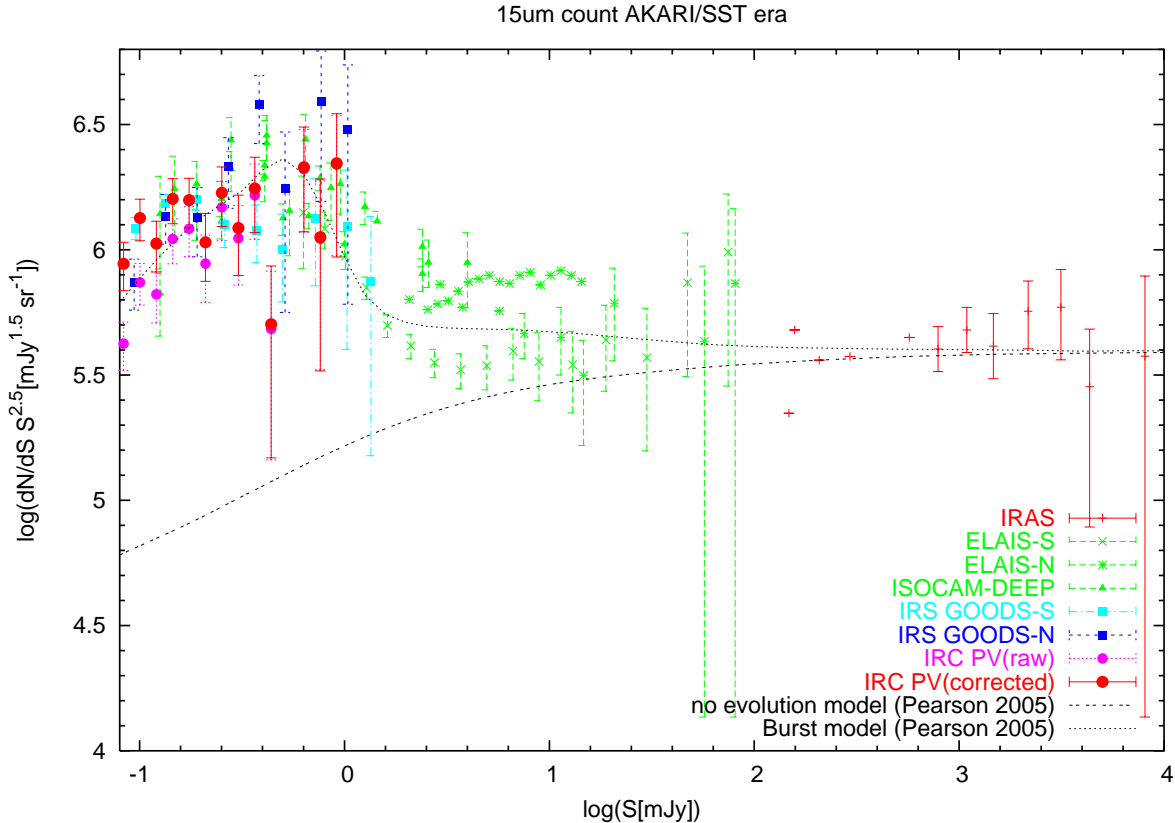}
   \end{center}
   \caption{
AKARI/IRC 15 $\mu$m differential source counts (both raw and
corrected for in-completeness) are plotted with 
the surveys carried out by  IRAS (shifted from 12  $\mu$m), ISO/ISOCAM and Spitzer/IRS results.
The source counts are normalized to a Euclidian  flat universe.
A no-evolution  model and the contemporary evolutionary model of  \citet{pearson-sst-iso-model}
 are also plotted for  reference.
Only flux density level above 50\% completeness limit of our results
is shown.

Spitzer/IRS results are from GOODS-N \citep{teplitz-16um-goods-n} and 
GOODS-S \citep{teplitz-16um-goods-s} survey.
Data points ISOCAM-DEEP include the ISO cluster lens survey 
(\cite{altieri-iso15-lens}), and 
the ISO Hubble Deep Field surveys (\cite{oliver97}; \cite{aussel99}). 
 ELAIS-S and ELAIS-N are from  the European Large-Area ISO
 Survey (\cite{gruppioni02}; \cite{serjeant00}).
IRAS counts are from \citet{rush93}.
}\label{fig:lognlogs_l15_wide}
\end{figure*}

Figure~\ref{fig:lognlogs_l15_wide} shows the 15 $\mu$m  number count results
of the  AKARI/IRC, ISO/ISOCAM and {\it Spitzer} IRS surveys. 
This work is currently the widest survey below 100 $\mu$Jy at this wavelength. 
The IRC, ISOCAM and IRS results are consistent with each other 
for sources brighter than 0.2 mJy, and the IRC results confirm
the existence of the excess compared to the no-evolution source count
models around 0.4mJy found in the ISO surveys \citep{elbaz-iso15}. 
This excess makes itself apparent at around the 1mJy level. However, there is still a large uncertainty
around the 1-10mJy flux range due to cosmological variance and calibration
issues with the ISOCAM data  \citep{vaisanen-isocal}. 
The results from the deep and wide field near- and mid- infrared surveys
by AKARI/IRC (the LS NEP DEEP and LS NEP WIDE surveys; \cite{matsuhara-lsnep}) 
will cover this flux range and will be able to settle the issue of the normalization of the bright source counts.

The large discrepancy between the actual source counts and predictions
by no-evolution models cannot be explained singly by 
K correction effects caused by the strong PAH emission from redshifted
galaxies and some new population is required to explain 
the excess (e.g.  \cite{cpp01}, \cite{xu03}, \cite{lagache03},
\cite{pozzi04}, and  \cite{{pearson-sst-iso-model}}). 
Moreover, evolutionary models which fit both the  ISO 15 $\mu$m
and  {\it Spitzer} 24 $\mu$m counts simultaneously 
have been already presented by \citet{pearson-sst-iso-model}.

\begin{figure*}
   \begin{center}
      \FigureFile(160mm,50mm){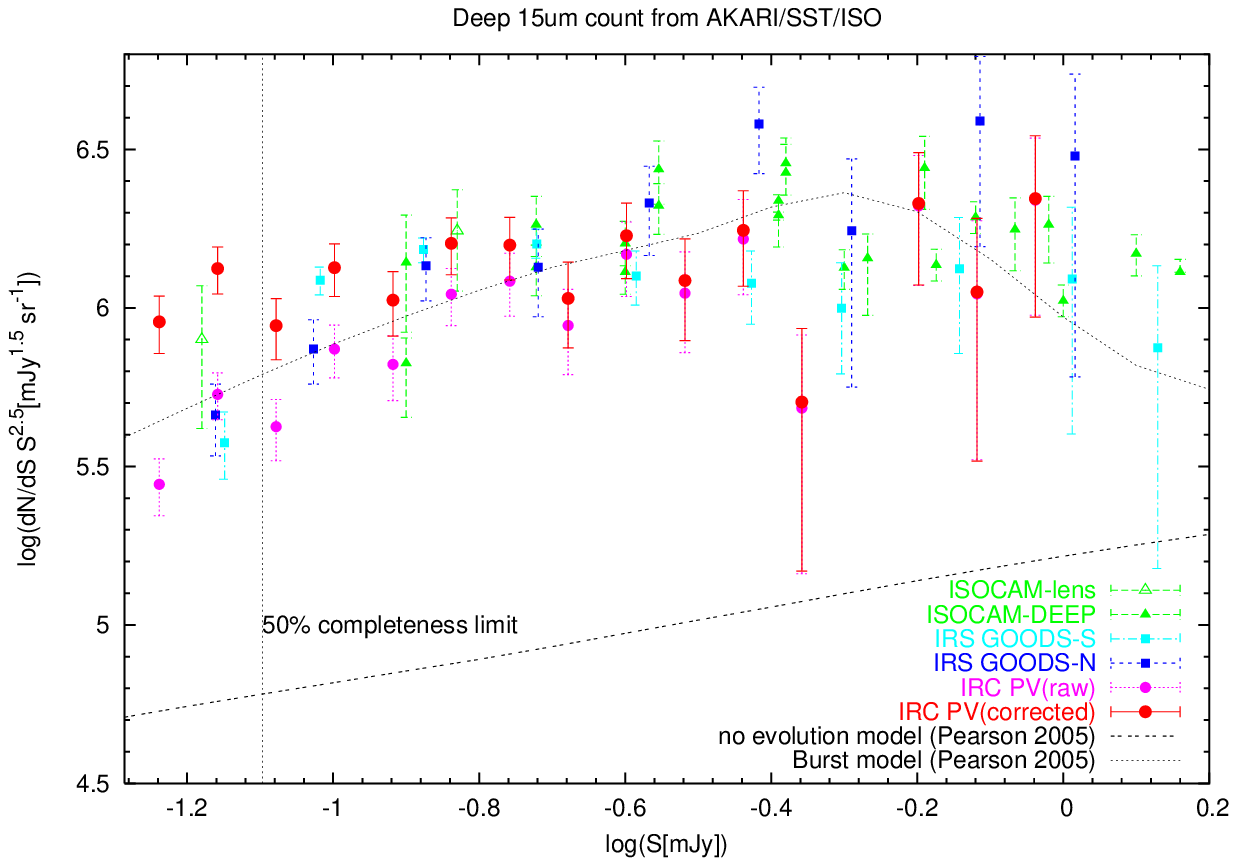}
   \end{center}
   \caption{
 The faint end of the 15 $\mu$m differential source  counts.
Only flux density level above 30\% completeness limit of our results
is shown. 
Correction for in-completeness may suffer from systematic bias at the
 fainter flux level, and 50\% completeness limit of our results is shown for reference.

Open triangles are data from the ISO gravitational lensing cluster
survey  of \citet{altieri-iso15-lens}.
Note that no correction for in-completeness is applied for Spitzer/IRS results.

 }\label{fig:lognlogs_l15_close}
\end{figure*}

Figure~\ref{fig:lognlogs_l15_close} shows the faint end of 
15 $\mu$m source counts.
The raw data of the IRC results at the faint end (below 100$\mu$Jy),
are consistent with the results from the {\it Spitzer} IRS 16$\mu$m PUI surveys in
the GOODS fields (\cite{teplitz-16um-goods-n}; \cite{teplitz-16um-goods-s}) in which 
no completeness correction is applied.
The source counts corrected for incompleteness, however, 
agree better with the faint results from the ISO/ISOCAM lensing survey by
\citet{altieri-iso15-lens}, and show a slightly higher number density 
 than the prediction from the model fit 
to the the ISOCAM results \citep{pearson-sst-iso-model}.
The faint end of the source counts, however,  may suffer from
systematic bias such as Eddington bias \citep{hogg-bias}.
An ultra deep survey by the AKARI/IRC or the {\it Spitzer}/IRS 16$\mu$m PUI
is essential to produce a definitive answer to the faint source counts.

\section{Summary}
The first near- and mid- infrared deep survey with the Infrared Camera
onboard AKARI has been successfully completed in the performance 
verification phase of the mission.
A total of 10 pointed observations have covered 86.0, 70.3, and 77.3 arcmin$^2$ 
area, with effective integration times of 4529, 4908, and 4417 seconds 
at 3, 7, and 15 $\mu$m,   have detected 955, 298 and 277 sources, 
respectively.
The 5 $\sigma$ detection limits of the survey are
6.0, 31.5 and 71.2$\mu$Jy and the 50 \% completeness limit 
are  24.0, 47.5, and  88.1$\mu$Jy at 3, 7, and 15 $\mu$m, respectively.
The observation is limited by source confusion at 3 $\mu$m.

We have confirmed the turnover of the  
differential source counts around 400$\mu$Jy discovered by 
the ISO 15 $\mu$m surveys. 
The faint end of our 15 $\mu$m raw source counts are consistent
with the Spitzer IRS results of \citet{teplitz-16um-goods-n} and 
the predictions from contemporary evolutionary models.

\bigskip

This work is based on the observations in the performance verification phase
of the AKARI satellite, which is developed and operated 
by ISAS/JAXA with collaboration with ESA, Universities and companies in 
Japan, Korea, UK, and The Netherlands.
The authors would like to thank all those who have been involved the AKARI project 
for many years.

\end{document}